\documentclass{article}

\usepackage{PRIMEarxiv}

\usepackage[utf8]{inputenc} 
\usepackage[T1]{fontenc}    
\usepackage{hyperref}       
\usepackage{url}            
\usepackage{booktabs}       
\usepackage{amsfonts}       
\usepackage{amsmath}
\usepackage{nicefrac}       
\usepackage{microtype}      
\usepackage{lipsum}
\usepackage{fancyhdr}       
\usepackage{graphicx}       
\graphicspath{{media/}}     

\title{Predicting treatment effects from observational studies using machine learning methods:  A simulation study}

\author{
  Bevan Smith \\
  Academic Development Unit \\
  University of the Witwatersrand \\
  Johannesburg\\
  \texttt{bevan.smith@wits.ac.za} \\
   \And
  Charles Chimedza \\
  School of Statistics and Actuarial Science \\
  University of the Witwatersrand \\
  Johannesburg\\
  \texttt{charles.chimedza@wits.ac.za} \\
}

\begin{document}
\maketitle

\begin{abstract}
Measuring treatment effects in observational studies is challenging because of confounding bias. Confounding occurs when a variable affects both the treatment and the outcome.   Traditional methods such as propensity score matching estimate treatment effects by conditioning on the confounders.  Recent literature has presented new methods that use machine learning to predict the counterfactuals in observational studies which then allow for estimating treatment effects.  These studies however, have been applied to real world data where the true treatment effects have not been known.  This study aimed to study the effectiveness of this counterfactual prediction method by simulating two main scenarios: with and without confounding.  Each type also included linear and non-linear relationships between input and output data.  The key item in the simulations was that we generated known true causal effects.  Linear regression, lasso regression and random forest models were used to predict the counterfactuals and treatment effects.  These were  compared these with the true treatment effect as well as a naive treatment effect.  The results show that the most important factor in whether this machine learning method performs well, is the degree of non-linearity in the data.   Surprisingly, for both non-confounding \textit{and} confounding, the machine learning models all performed well on the linear dataset.  However, when non-linearity was introduced, the models performed very poorly.  Therefore under the conditions of this simulation study, the machine learning method performs well under conditions of linearity, even if confounding is present, but at this stage should not be trusted when non-linearity is introduced.
\end{abstract}


\section{Introduction}
\label{sec1}

Estimating treatment effects in observational studies is a key component in determining if an academic intervention is effective.  Traditional ways of estimating treatment effects in observational studies include matching, propensity score matching (PSM) \cite{Ye2009,Stuart2010},  and inverse probability of treatment weighting (IPTW) \cite{Stuart2010}.  These aim to measure treatment effects by creating a \textit{counterfactual} by matching observations in treated and control groups.  However, this causes the dataset to shrink due to the challenge of finding matching cases that balance treated and control groups.  

Recent work has applied a method using machine learning to measure treatment effects in observational studies, \textit{not by balancing} the control and treated datasets to obtain counterfactuals, but by predicting the counterfactuals \cite{Beemer2017, Beemer2018, BevanSmith2020}. This method has the benefit of not needing to balance the groups, thereby \textit{preserving the full dataset} and not losing any observations.

The problem is that although this counterfactual prediction method shows promise, the above-mentioned studies \cite{Beemer2017,Beemer2018,BevanSmith2020} were carried out on real-life data where the true treatment effect was not known.  Smith et al. \cite{BevanSmith2020} attempted to validate the counterfactual prediction method by comparing it with PSM estimations; however the ground truth average treatment effect (ATE) was still not known.  

The main aim of this study therefore was to estimate how well this counterfactual prediction method works by performing simulation studies where the true treatment effect, $ATE_{true}$, is known.  We simulated various scenarios that included datasets with no confounding, with confounding, and with linear and non-linear relationships in the data.  The models used included linear regression, lasso regression and random forest.   

The main finding of this study is that the counterfactual prediction method using machine learning to predict counterfactuals, works well as long as the data is linear, regardless of whether there is no confounding or single feature confounding.  This method breaks down quite substantially when non-linearity is introduced.

\section{Observational studies and treatment effects}
\label{sec2}
  Traditional ways of measuring treatment (causal) effects are either to carry out a randomized control trial (RCT), the gold standard of experimental studies \cite{Austin2011}, or controlling for observed confounders in observational studies \cite{Cochran1974ControllingBI}.  Although the benefits of a randomized trial are evident, we often only have access to observational data, which generally produces biased treatment effects due to self-selection confounding \cite{Austin2011,Morgan2007}. 

Treatment effects can be measured via the average treatment effect, ATE, seen in Equation \ref{eq:1} which is the difference between the average outcome of the treated and control groups.

\begin{equation}
	ATE = E(Y_t) - E(Y_c)
	\label{eq:1}
\end{equation}

However, in observational studies, ATE is generally not the true treatment effect, but a naive effect containing bias due to confounding. Equation \ref{eq:1} will therefore in general not estimate the true treatment effect.    The true ATE in observational studies can be computed using Equation \ref{eq:2} \cite{Morgan2007},

\begin{equation}
	ATE = \pi(E(Y_t) - E(\hat{Y_t})) +(1-\pi)(E(\hat{Y_c})- E(Y_c))
	\label{eq:2}
\end{equation}

where $\hat{Y_t}$ and $\hat{Y_c}$ are the counterfactual outcomes for the treated and control groups respectively and ${Y_t}$ and ${Y_c}$ are the actual outcomes of the treated and control groups respectively.  The quantity $\pi$ refers to the fraction of observations receiving the treatment, the participation rate. An example in a higher education setting might be that we are aiming to estimate the treatment effect of online videos on final grades of a course.  The treated group could be those students that watched supplementary online videos and the control group would be those that didn't.  Therefore the treatment is the watching of videos. It is important to note that the counterfactual outcomes cannot be measured (this is the fundamental problem of causal inference), and the great challenge here is to find a method that can predict the counterfactuals.

\section{Confounding bias in  observational studies}\label{sec4}

\begin{figure}[htbp!]
	\centering
	\includegraphics[width=50mm]{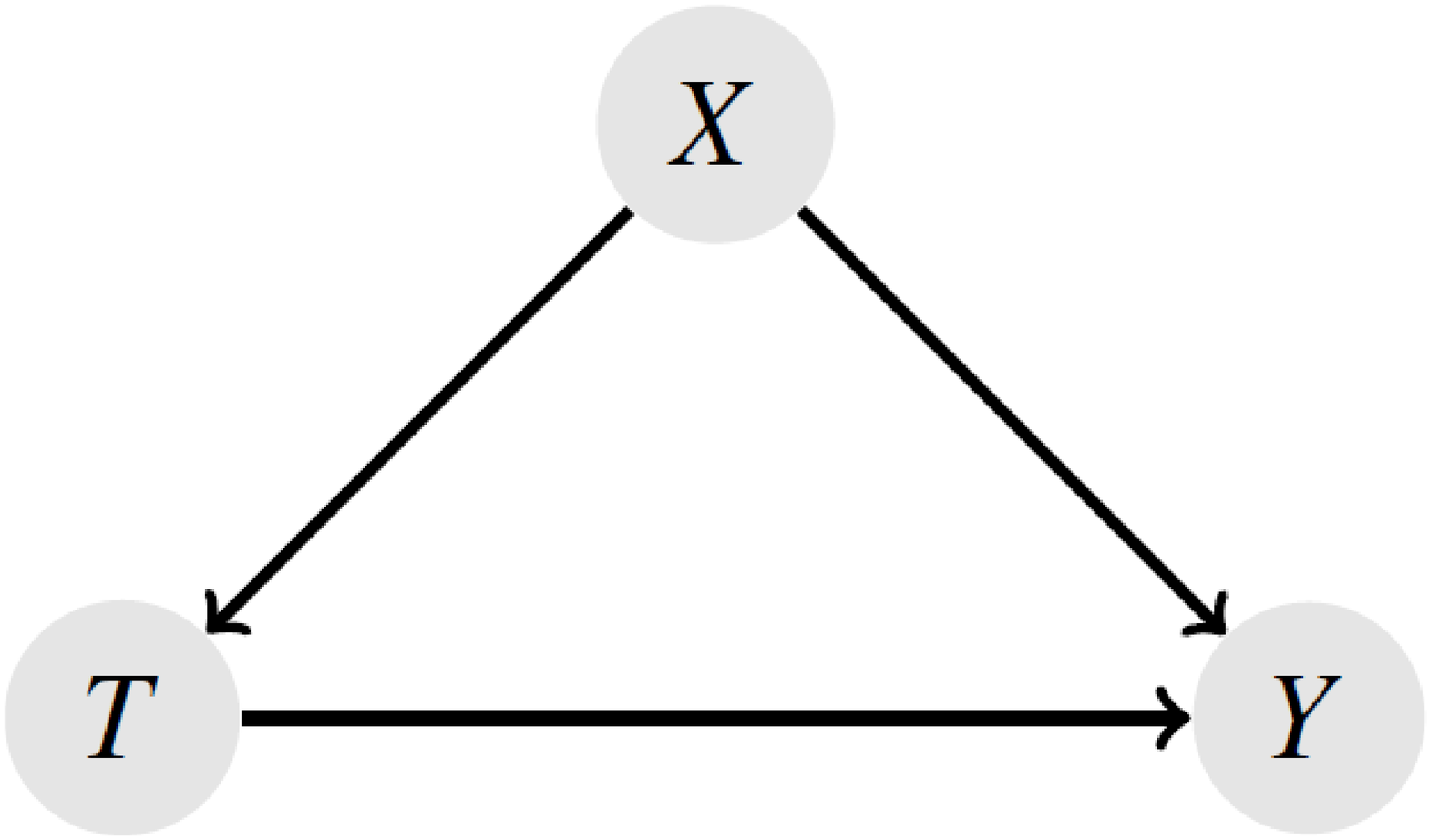}
	\caption{Directed acyclic graph showing confounding.}
	\label{Fig1}
\end{figure}

In observational studies such as ones we come across when students self-select academic interventions, bias due to confounding occurs when both the outcome \textit{Y} and the treatment \textit{T} are caused by a common parent \textit{X} \cite{BradyNeal2020}, shown in the directed acyclic graph (DAG) in  Figure \ref{Fig1}.  An example is where students self-select an intervention (\textit{T}), say extra videos or tutorials,  based on their current grades (\textit{X}), which also affect the outcome (\textit{Y}).  The problem is that \textit{T} possesses inherent bias that does not result in the true ATE when applying Equation \ref{eq:1}.

\begin{figure}[htbp!]
	\centering
	\includegraphics[width=50mm]{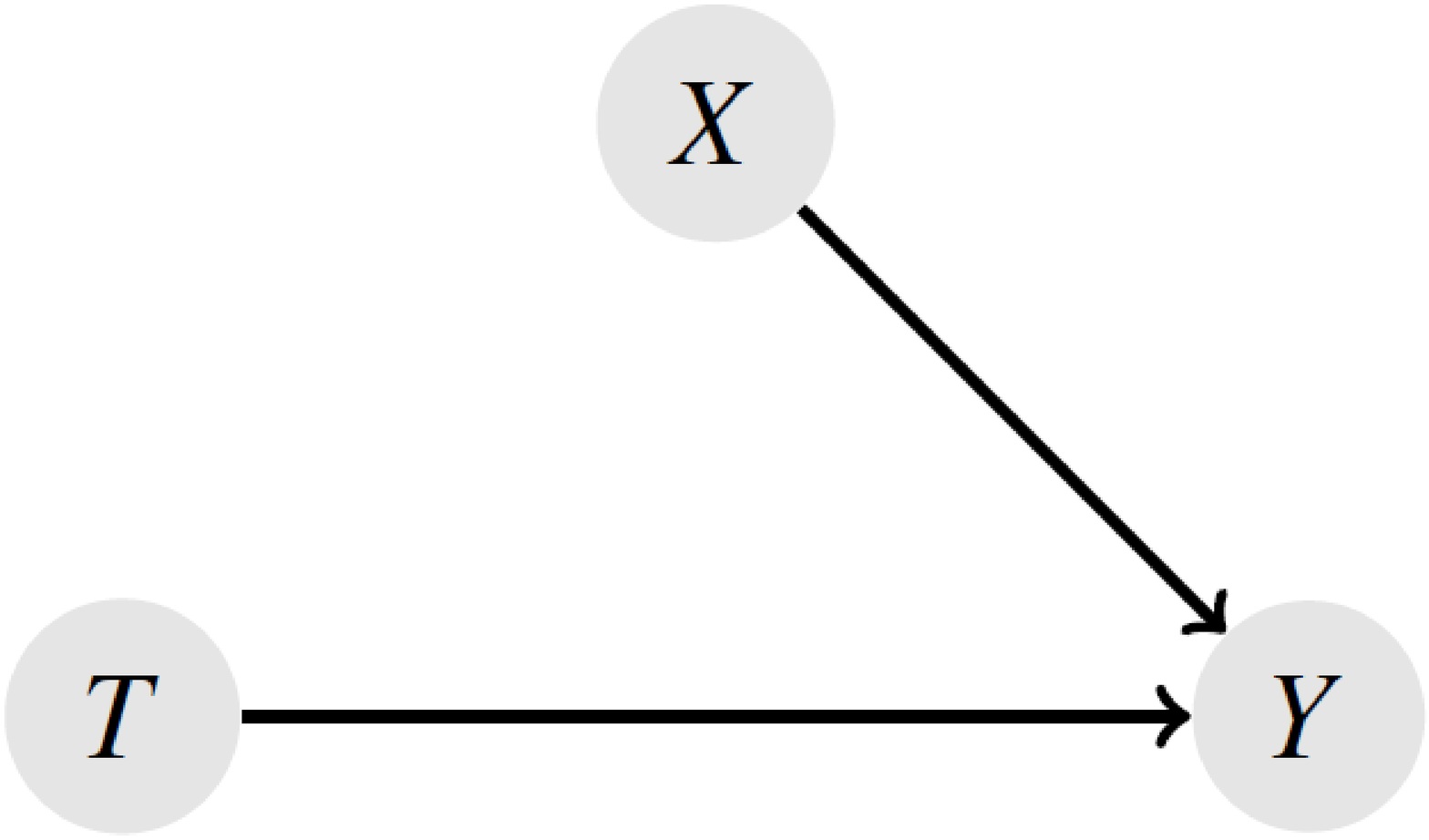}
	\caption{Directed acyclic graph showing no confounding.}
	\label{Fig2}
\end{figure}

Ideally, we want \textit{T} to be independent of \textit{X} so that the unbiased treatment effect \cite{Austin2011} of \textit{T} on \textit{Y} can be measured, shown in Figure \ref{Fig2}.  The scenario in Figure \ref{Fig2} can take place if we randomize which students take the treatment and thereby destroy any causal relationship with \textit{X}.  This is however not always possible due to ethical reasons; we cannot refuse any students access to an academic intervention.

\section{Counterfactual prediction method}
\label{sec3}
The counterfactual prediction method introduced earlier is described here \cite{Beemer2017,Beemer2018,BevanSmith2020}.  Predictions are carried out via machine learning models. For a dataset where a treatment was administered, the method for predicting the counterfactuals and computing the treatment effect is as follows:

\begin{enumerate}
	\item {Split the entire dataset into treated and control groups.}
	\item{Train a machine learning model $M_t$ on the treated group and a model $M_c$ on the control group.}
	\item{Predict $\hat{Y_t}$ (treated group counterfactual outcome) by feeding the treated group input features, $X_t$, into the control group model  $M_c$.  Predict $\hat{Y_c}$ (control group counterfactual outcome) by feeding the control group input features, $X_c$, into the treatment  model  $M_t$. }
	\item{Estimate the true treatment effect using Equation \ref{eq:2}.}
\end{enumerate}

This method therefore uses machine learning to predict counterfactuals by predicting what the treated group would have obtained, had they received the control and what the control group would have obtained, had they received the treatment.  The assumption with this method is that the counterfactual of say, the treated group, has the characteristics of the control group and the counterfactual of the control group, has the characteristics of the treated group.  This is because we feed a group's input features into its' counterfactual model.

\section{Simulations }
Various simulations testing the counterfactual prediction method were generated based on two main scenarios: no confounding and confounding.   For both scenarios we looked at cases where the relationship between input and output features was linear and non-linear.   The aim was to ascertain how well the counterfactual prediction method works under various scenarios and how well different machine learning models perform when used as the predictive model.  The models being compared were linear regression, lasso models and random forest.  

\subsection{Dataset features}
Our generated dataset, shown in Table \ref{TabData}, comprised five input features (including a treatment indicator) and a target output feature.  The dataset mimics data from a university course, where the input features represent grades and demographic features of students, and the output is a final course grade.  Table \ref{TabData} shows the features and how they were generated.  The two grades features describe grades where the mean values are 50\% and 45\% respectively.  The Age feature describes the age distribution of first year students.  Gender is slightly skewed to one side with a probability of 0.6. For the initial randomized simulation, Treatment is randomly assigned to each example, thereby being independent of any input features and simulating a randomized trial.  Treatment could be any intervention such as extra tutorials or videos etc.  The sample size generated for each simulation was 1000 and for each scenario, we carried out 1000 simulations. 

\begin{table}[htbp!]
	\centering
	\caption{Description of generated data.}
	
	\begin{tabular}{lll}
		\hline
		\textbf{Name} & \textbf{Description} & \textbf{Distribution}\\
		\hline
		x1 & Grades  &  Normal, $\mu$ = 50, sd = 5 \\
		x2 & Age & Normal, $\mu$ = 20, sd = 2 (minimum of 18) \\
		x3 & Grades & Normal, $\mu$ = 45, sd = 6 \\
		x4 & Gender & Binomial, prob = 0.6 \\
		T & Treatment & Binomial, prob = 0.5\\
		\hline
	\end{tabular}
	\label{TabData}
\end{table}
\newpage
The following simulations were run which are elaborated on in the subsequent sections.

\begin{enumerate}
	\item{Simulation 1a: No confounding;  linear dataset}
	\item{Simulation 1b: No confounding; non-linear (squared)}
	\item{Simulation 1c: No confounding;  non-linear (cubed)}
	\item{Simulation 2a: Confounding; linear dataset}
	\item{Simulation 2b: Confounding, non-linear (squared)}
	\item{Simulation 2c: Confounding; non-linear (cubed)}
\end{enumerate}

\subsection{Simulation 1: no confounding; linear relationship}
\label{no_confounding}
We first simulated a randomized control trial where the treatment was independent of the input features, i.e. there was no confounding (as per Figure \ref{Fig2}). For the no confounding and linear simulation, an output vector \textit{Y} was generated from the input features plus a Gaussian error of zero mean and standard deviation of 1, by assuming a linear relationship as per Equation \ref{eq:3}:

\begin{equation}
	Y = \beta_0 + \beta_1 x_1 + \beta_2 x_2 + \beta_3 x_3 + \beta_4 x_4 + ATE_{true} T  + \mathcal{N} (0,1)
	\label{eq:3}
\end{equation}	

\vspace{0.3cm}

Arbitrary values for the coefficients were chosen as follows: $\beta_0 = 0.5$, $\beta_1 = 0.7$, $\beta_2 = 0.5$, $\beta_3 = 0.5$, $\beta_4$ = 0.7.  

\subsubsection{No confounding; non-linear relationship (squared)}
\label{non-linear2}
We further simulated no confounding and non-linearity by making the output \textit{Y} have a squared non-linear relationship with $x_1$ as shown in Equation \ref{eq:33}.  The remainder of the features all had a linear relationship with \textit{Y}.

\begin{equation}
	Y = \beta_0 + \beta_1 x_1^2 + \beta_2 x_2 + \beta_3 x_3 + \beta_4 x_4 + ATE_{true} T  + \mathcal{N} (0,1)
	\label{eq:33}
\end{equation}

\subsubsection{No confounding; non-linear relationship (cubed)}
\label{non-linear3}
We next simulated another non-linear scenario by making the output \textit{Y} have a cubic non-linear relationship with $x_1$ as shown in Equation \ref{eq:43}.  The remainder of the features all had a linear relationship with \textit{Y}.

\begin{equation}
	Y = \beta_0 + \beta_1 x_1^3 + \beta_2 x_2 + \beta_3 x_3 + \beta_4 x_4 + ATE_{true} T  + \mathcal{N} (0,1)
	\label{eq:43}
\end{equation}

\subsubsection{Ranges of \texorpdfstring{$ATE_{true}$} aand \texorpdfstring{$\pi$}  vvalues}
A further aim was to study a range of true ATE values as well as a range of treatment participation ($\pi$) rates.  $ATE_{true}$ was varied as follows: [-10, -5, 0.1, 5, 10] and $\pi$ was varied as follows: [0.1, 0.3, 0.5, 0.7, 0.9].  Note that the third $ATE_{true}$ value was 0.1 which was used instead of zero to prevent dividing by zero. Equation \ref{eq:2} was used to compute the ATE.  Here we rename it $ATE_{sim}$ (Equation \ref{eq:22}) to indicate it was computed from the simulation.
\begin{equation}
	ATE_{sim} = \pi(E(Y_t) - E(\hat{Y_t})) +(1-\pi)(E(\hat{Y_c})- E(Y_c))
	\label{eq:22}
\end{equation}

We therefore computed $ATE_{sim}$ using the machine learning counterfactual prediction method for a range of true treatment effects and a range of participation rates.   Once $ATE_{sim}$ was calculated for a given $ATE_{true}$ and $\pi$, the absolute error was calculated as per Equation \ref{eq:4}.  The number of simulations was 1000.

\begin{equation}
error = abs(100*\frac{ATE_{sim} - ATE_{true}}{ATE_{true}})
\label{eq:4}
\end{equation}

\begin{table}[htbp!]
	
	\label{Algo1}
	\begin{tabular}{l}
		\hline
		Pseudocode 1 for each run of a no confounding simulation \\
		\hline
		1. Generate data: $x_1$ to $x_4$, \textit{T} and \textit{Y} as per Table \ref{TabData} and Equations \ref{eq:3}, \ref{eq:33} and \ref{eq:43},\\ depending on type of simulation.\\
		2. Split dataset into treated and control groups.\\
		3. Train the three models (linear model $M_t$, lasso $L_t$, random forest $R_t$)  on treated group; \\train the three models (linear model $M_c$, lasso $L_c$, random forest $R_c$) on control group.\\
		4. Feed treated group input features $x_1$ to $x_4$ into linear model $M_t$, lasso $L_t$, random forest $R_t$ \\to predict treated group counterfactual $\hat{Y_t}$ for each model; \\ feed control group $x_1$ to $x_4$ into $M_t$, $L_t$ and $R_t$ to predict control group counterfactual $\hat{Y_c}$ \\for each model.\\
		5. Compute $ATE_{sim}$ for the three models, as per Equation \ref{eq:2} and store.\\
		6. Compute absolute percentage error for the three models as per Equation \ref{eq:4} and store.\\ \hline
		
	\end{tabular}
\end{table}

\subsection{Simulation 2: confounding}
We next introduce confounding via a single feature, $x_3$, to simulate an observational study.  That is, we make $x_3$ a parent of both \textit{T} and \textit{Y}, as shown in Figure \ref{Fig1}.  Treatment is now dependent on a single observable feature. The confounding is created as per Equation \ref{eq:5}: if any student obtained below 41\% for their $x_3$ grade, they would attend the intervention, i.e. receive treatment, and hence will be encoded as 1.  If any student obtains above 49\% they would not attend the intervention, and will be encoded as 0.  Between 41\% and 49\% we generate a probability of  0.5 based on a normal distribution, indicating that in this region there is a 50\% probability of the students receiving the treatment.  We generated 1000 simulations in the same manner as Simulation 1 based on the same pseudocode.

\begin{equation}
T = \begin{cases}
1 & \text{if $x_3$ $<$ 49\%}\\
0 & \text{if $x_3$ $>$ 41\%}\\
0.5 ~probability & \text{otherwise}
\end{cases}
\label{eq:5}
\end{equation}

We run confounding simulations in exactly the same way as that for no confounding, generating data using Equations \ref{eq:3}, \ref{eq:33} and \ref{eq:43} as discussed in Section \ref{no_confounding} as well as ranges of $ATE_{true}$ and $\pi$.  The only difference is the feature T described above where confounding is introduced (Equation \ref{eq:5}).

\section{Models used}
\label{sec5}
This study made use of three models to carry out the counterfactual predictions in the simulations:  linear regression, lasso regression and random forest.  Linear regression was used as a base model. Lasso regression and random forest were used to see if they perform better than linear regression when non-linearity and confounding are present in the data. 

The linear regression model is a basic parametric model using ordinary least squares to estimate the coefficients.  The lasso model \cite{Hastie2009} is a regularized regression model that is able to shrink to zero correlated features by introducing a penalty on the coefficients as follows:

\begin{equation}
\lambda \sum_{j=1}^{p} \lvert \beta_j\lvert
\end{equation}

where $\lambda$ is a penalty, \textit{p} is the number of features, and $\beta$ refers to the coefficients in the parametric model.   When training the lasso models, 10-fold cross validation was employed to obtain the $\lambda$ values that minimize the mean-squared-error during training. Once the $\lambda$ values were obtained, they were utilized in the lasso models to carry out predictions. 

The third model random forest, is an ensemble method that generates multiple decision trees and averages the outcome for regression problems such as the one in this study.  It is a non-parametric machine learning model that generates each tree by carrying out bootstrap aggregating on the observations and random selection of a subset of the features.  This allows for the trees in the ensemble to be decorrelated and produce superior results when compared with single decision trees. They also are able to handle non-linearity well due to their non-parametric nature \cite{Hastie2009}.

\section{Software}
\label{software}
The simulations were run in the RStudio integrated development environment using the \textit{glmnet} package for lasso and the \textit{caret} package for random forest.  

\section{Results}
The results of the no confounding simulations are presented first, followed by confounding.  The aim of the various simulations was to compare simulated ATE computations, $ATE_{sim}$, with $ATE_{true}$, under different scenarios.  For all the scenarios, four ATE quantities are compared with $ATE_{true}$:  $ATE_{naive}$ (Equation \ref{eq:1}) and $ATE_{sim}$ (Equation \ref{eq:22}) for linear regression, lasso regression and random forest.  The absolute errors of each simulation were also computed as per Equation \ref{eq:4} as well as the average error over all the ranges of parameters.

\subsection{No confounding; linear relationship}
$ATE_{sim}$ and error results for the no confounding simulation with linear relationship, are presented in Tables \ref{TabNCL1} and \ref{TabNCL2}, respectively.  In Table \ref{TabNCL1}, Naive, LM, Lasso and RF refer to the $ATE_{sim}$ computations for the naive, linear regression model, lasso model and random forest respectively.  Furthermore, E(Naive), E(LM), E(Lasso) and E(RF) refer to the errors calculated for the different methods, using Equation \ref{eq:4}.  This is the same for all subsequent tables.

The results show that regardless of the method used, whether naive or machine learning models, all ATE predictions were relatively accurate, with LM performing the best with an average error of 0.52\% (Table \ref{TabNCL2}).  Random forest performed worst with an average error of 2.04\%. In terms of how well the methods perform as $ATE_{true}$ is varied, the highest errors occur at $ATE_{true}$ = 0.1.  This can be attributed to dividing by a fraction that magnifies the errors.  Therefore for a scenario with no confounding and a linear relationship between input and output data, all methods, whether naive or machine learning predictions, can provide a good approximation of the true treatment effect, with linear regression performing the best.


\begin{table}[htbp!]
	\centering
	\caption{$ATE_{sim}$ results for no confounding, linear simulation.}
	\begin{tabular}{cccccccccc}
		$ATE_{true}$ & $\pi$ & Naive  & E(Naive) & LM & E(LM) & Lasso & E(Lasso)  & RF & E(RF)  \\ \hline
			 & 0.1 & -10.02 & 0.16  & -9.99  & 0.09 & -9.96  & 0.38  & -10.00 & 0.04  \\
                     & 0.3 & -10.00 & 0.02  & -10.00 & 0.00 & -9.98  & 0.24  & -10.01 & 0.06  \\
                   -10  & 0.5 & -10.01 & 0.14  & -10.00 & 0.01 & -9.99  & 0.15  & -10.00 & 0.00  \\
                     & 0.7 & -10.00 & 0.00  & -10.00 & 0.02 & -9.99  & 0.08  & -9.99  & 0.07  \\
                     & 0.9 & -10.02 & 0.19  & -10.01 & 0.05 & -10.00 & 0.02  & -9.99  & 0.08  \\ \hline
 & 0.1 & -5.02  & 0.40  & -5.00  & 0.05 & -4.98  & 0.31  & -5.02  & 0.35  \\
                     & 0.3 & -4.99  & 0.25  & -5.00  & 0.03 & -4.99  & 0.27  & -5.01  & 0.14  \\
                   -5  & 0.5 & -5.00  & 0.02  & -5.00  & 0.07 & -4.99  & 0.26  & -5.00  & 0.07  \\
                     & 0.7 & -5.01  & 0.20  & -5.00  & 0.01 & -5.00  & 0.08  & -5.00  & 0.09  \\
                     & 0.9 & -5.02  & 0.39  & -5.00  & 0.02 & -5.00  & 0.01  & -4.98  & 0.34  \\ \hline
 & 0.1 & 0.11   & 12.27 & 0.10   & 3.57 & 0.11   & 10.15 & 0.09   & 7.89  \\
                     & 0.3 & 0.09   & 11.75 & 0.10   & 4.25 & 0.10   & 0.34  & 0.09   & 14.69 \\
                  0.1   & 0.5 & 0.10   & 1.97  & 0.10   & 0.18 & 0.10   & 2.53  & 0.10   & 0.61  \\
                     & 0.7 & 0.11   & 6.65  & 0.10   & 3.03 & 0.10   & 3.55  & 0.11   & 14.02 \\
                     & 0.9 & 0.10   & 0.75  & 0.10   & 1.29 & 0.10   & 0.99  & 0.11   & 10.99 \\ \hline
   & 0.1 & 4.99   & 0.27  & 5.00   & 0.02 & 4.99   & 0.11  & 4.98   & 0.47  \\
                     & 0.3 & 5.01   & 0.25  & 5.00   & 0.02 & 4.99   & 0.13  & 4.99   & 0.11  \\
                   5  & 0.5 & 4.98   & 0.32  & 5.00   & 0.04 & 4.99   & 0.14  & 5.00   & 0.06  \\
                     & 0.7 & 5.00   & 0.01  & 5.00   & 0.02 & 5.00   & 0.09  & 5.01   & 0.16  \\
                     & 0.9 & 5.01   & 0.25  & 4.99   & 0.15 & 4.99   & 0.17  & 5.01   & 0.19  \\ \hline
  & 0.1 & 10.00  & 0.02  & 10.00  & 0.01 & 9.98   & 0.18  & 9.98   & 0.17  \\
                     & 0.3 & 9.99   & 0.12  & 10.00  & 0.00 & 9.98   & 0.17  & 9.99   & 0.11  \\
                 10    & 0.5 & 9.99   & 0.07  & 10.00  & 0.02 & 9.99   & 0.14  & 10.00  & 0.02  \\
                     & 0.7 & 9.99   & 0.05  & 10.00  & 0.02 & 9.99   & 0.10  & 10.01  & 0.06  \\
                     & 0.9 & 10.00  & 0.01  & 10.00  & 0.02 & 10.00  & 0.05  & 10.01  & 0.15 \\ \hline
	\end{tabular}
		\label{TabNCL1}
\end{table}

\begin{table}[htbp!]
    \centering
    \caption{Mean absolute percentage errors for simulations with no confounding and linear  relationship.}
    \begin{tabular}{cccc}
         \textbf{Naive}& \textbf{LM} & \textbf{Lasso} & \textbf{RF}  \\ \hline
         1.46 & 0.52 & 0.82 & 2.04 \\ \hline
    \end{tabular}
     \label{TabNCL2}
\end{table}

\subsection{No confounding; non-linear relationship (squared)}
In this no confounding simulation, non-linearity ($x_1^2)$ was introduced as per Equation \ref{eq:33}.  The results are shown in Tables \ref{TabNCNL2} and \ref{TabErrorNCNL2}.  An overall view of Table \ref{TabNCNL2} shows that the non-linearity has introduced less uniform and more unstable $ATE_{sim}$ predictions.  The errors are also considerably larger than for the linear simulation.  Looking at Table \ref{TabErrorNCNL2} we can see that just by introducing squared non-linearity on one of the features, the errors are approximately two orders of magnitude larger than for the linear simulation.  Again, on average, the linear regression model (LM) performed the best with 33.3\% absolute error.  We also now see that lasso has performed the worst (371.7\%) and random forest (RF) is second to LM.  These results suggest that when non-linearity exists in the dataset, these methods begin to break down.

\begin{table}[htbp!]
	\centering

	\caption{$ATE_{sim}$ results for no confounding, non-linear (squared) simulation.}
	\begin{tabular}{cccccccccc}
		
		$ATE_{true}$ & $\pi$ & Naive  & E(Naive) & LM & E(LM) & Lasso & E(Lasso)  & RF & E(RF)  \\ \hline
		 & 0.1 & -9.16  & 8.39    & -10.46 & 4.62   & -5.65 & 43.47   & -10.64 & 6.44   \\
                     & 0.3 & -11.75 & 17.46   & -10.04 & 0.41   & -7.60 & 24.04   & -10.22 & 2.24   \\
                   -10  & 0.5 & -9.62  & 3.79    & -9.96  & 0.43   & -8.62 & 13.78   & -10.03 & 0.29   \\
                     & 0.7 & -11.50 & 14.99   & -9.93  & 0.69   & -9.63 & 3.66    & -9.83  & 1.71   \\
                     & 0.9 & -9.58  & 4.20    & -9.74  & 2.56   & -9.75 & 2.52    & -9.25  & 7.51   \\ \hline
& 0.1 & -4.86  & 2.82    & -5.36  & 7.12   & -0.86 & 82.76   & -5.67  & 13.43  \\
                     & 0.3 & -4.53  & 9.37    & -5.12  & 2.32   & -2.26 & 54.90   & -5.19  & 3.76   \\
                   -5  & 0.5 & -5.10  & 2.00    & -4.91  & 1.86   & -3.49 & 30.22   & -5.05  & 1.07   \\
                     & 0.7 & -5.16  & 3.24    & -4.92  & 1.63   & -4.61 & 7.81    & -4.90  & 2.04   \\
                     & 0.9 & -5.46  & 9.29    & -4.64  & 7.14   & -4.51 & 9.88    & -4.35  & 12.98  \\ \hline
 & 0.1 & -1.57  & 1674.06 & -0.22  & 315.73 & 4.35  & 4253.46 & -0.70  & 796.08 \\
                     & 0.3 & -1.61  & 1710.30 & 0.00   & 101.04 & 2.37  & 2269.62 & -0.10  & 199.40 \\
                   0.1  & 0.5 & -0.05  & 150.63  & 0.11   & 5.16   & 1.38  & 1283.76 & 0.08   & 23.51  \\
                     & 0.7 & 0.75   & 650.50  & 0.19   & 93.87  & 0.88  & 780.07  & 0.26   & 161.65 \\
                     & 0.9 & -0.74  & 839.71  & 0.37   & 266.39 & 0.28  & 180.46  & 0.56   & 455.55 \\ \hline
  & 0.1 & 5.69   & 13.71   & 4.78   & 4.43   & 9.13  & 82.67   & 4.44   & 11.26  \\
                     & 0.3 & 5.15   & 3.04    & 4.91   & 1.77   & 7.32  & 46.49   & 4.86   & 2.79   \\
                    5 & 0.5 & 3.64   & 27.23   & 5.03   & 0.62   & 6.25  & 24.96   & 4.97   & 0.58   \\
                     & 0.7 & 5.06   & 1.24    & 5.01   & 0.27   & 5.36  & 7.18    & 5.13   & 2.60   \\
                     & 0.9 & 4.13   & 17.45   & 5.26   & 5.27   & 5.11  & 2.11    & 5.76   & 15.17  \\ \hline
  & 0.1 & 8.77   & 12.35   & 9.72   & 2.75   & 13.66 & 36.65   & 9.29   & 7.13   \\
                     & 0.3 & 9.99   & 0.11    & 9.99   & 0.11   & 12.58 & 25.81   & 9.83   & 1.73   \\
                   10  & 0.5 & 11.35  & 13.47   & 10.00  & 0.01   & 11.53 & 15.30   & 10.00  & 0.03   \\
                     & 0.7 & 10.24  & 2.43    & 10.16  & 1.64   & 10.78 & 7.78    & 10.21  & 2.15   \\
                     & 0.9 & 10.65  & 6.52    & 10.37  & 3.73   & 10.23 & 2.28    & 10.58  & 5.76 \\ \hline
	\end{tabular}
		\label{TabNCNL2}
\end{table}

\begin{table}[htbp!]
    \centering
    \caption{Mean errors for simulation with no confounding and non-linear (squared)  relationship.}
    \begin{tabular}{cccc}
         \textbf{Naive}& \textbf{LM} & \textbf{Lasso} & \textbf{RF}  \\ \hline
         207.9 & 33.3 & 371.7 & 69.5 \\ \hline
    \end{tabular}
     \label{TabErrorNCNL2}
\end{table}

\subsection{No confounding; non-linear relationship (cubed)}
In this no confounding simulation, non-linearity ($x_1^3)$ was introduced as per Equation \ref{eq:43}. The results are presented in  Tables \ref{TabConfCu} and \ref{TabErrorConfCu}.  For the squared non-linearity of the previous section, all methods began to perform poorly.  For cubed non-linearity however, the results are egregiously high and this method has completely broken down.

\begin{table}[htbp!]
	\centering
	\caption{$ATE_{sim}$ results for no confounding and cubic non-linear simulation.}
	\begin{tabular}{ccllllllll}
			$ATE_{true}$ & $\pi$ & Naive  & E(Naive) & LM & E(LM) & Lasso & E(Lasso)  & RF & E(RF)  \\ \hline
 & 0.1 & -18.9  & 89.4     & -56.1 & 460.5   & 379.7 & 3896.6   & -111.8 & 1018.3   \\
                     & 0.3 & -31.3  & 213.0    & -19.1 & 91.4    & 280.6 & 2905.5   & -40.1  & 300.7    \\
                 -10    & 0.5 & -147.9 & 1379.1   & -13.1 & 30.8    & 98.6  & 1086.2   & -9.3   & 6.9      \\
                     & 0.7 & 10.6   & 206.4    & 1.9   & 118.5   & 53.9  & 639.4    & 14.5   & 245.4    \\
                     & 0.9 & 124.0  & 1340.1   & 55.6  & 655.6   & 75.8  & 857.7    & 99.4   & 1093.5   \\ \hline
  & 0.1 & -88.8  & 1676.3   & -54.6 & 991.6   & 394.2 & 7983.7   & -111.4 & 2127.3   \\
                     & 0.3 & 36.1   & 821.6    & -24.6 & 392.5   & 263.8 & 5375.8   & -30.5  & 509.4    \\
                   -5  & 0.5 & 56.4   & 1227.7   & 7.0   & 239.6   & 177.9 & 3657.1   & -4.9   & 1.5      \\
                     & 0.7 & -20.7  & 313.5    & -0.9  & 82.2    & 71.8  & 1535.8   & 22.0   & 539.4    \\
                     & 0.9 & 102.3  & 2146.4   & 33.1  & 762.9   & 26.9  & 637.2    & 87.9   & 1857.4   \\ \hline
 & 0.1 & -127.6 & 127673.5 & -58.1 & 58242.2 & 403.2 & 403079.3 & -114.0 & 114096.4 \\
                     & 0.3 & -43.7  & 43799.8  & -15.8 & 15890.0 & 248.9 & 248761.4 & -33.2  & 33284.1  \\
                   0.1  & 0.5 & -3.6   & 3656.1   & -2.5  & 2639.7  & 161.6 & 161523.8 & -4.8   & 4888.4   \\
                     & 0.7 & -56.3  & 56404.3  & 14.8  & 14700.0 & 58.6  & 58523.0  & 25.6   & 25548.6  \\
                     & 0.9 & -4.1   & 4243.3   & 44.4  & 44328.7 & 35.2  & 35117.6  & 102.4  & 102272.8 \\ \hline
   & 0.1 & 60.5   & 1109.3   & -53.1 & 1162.4  & 415.1 & 8202.2   & -107.0 & 2239.8   \\
                     & 0.3 & -24.8  & 595.6    & -15.2 & 404.8   & 245.2 & 4803.6   & -25.4  & 608.9    \\
                  5   & 0.5 & 63.3   & 1165.8   & 15.9  & 218.8   & 152.1 & 2941.7   & 5.5    & 10.3     \\
                     & 0.7 & -6.8   & 235.3    & 13.5  & 170.4   & 51.5  & 930.6    & 28.9   & 478.3    \\
                     & 0.9 & -77.4  & 1648.0   & 56.1  & 1021.5  & 66.1  & 1221.4   & 110.4  & 2108.4   \\ \hline
  & 0.1 & -6.6   & 165.6    & -30.7 & 407.0   & 428.2 & 4182.3   & -95.2  & 1052.2   \\
                     & 0.3 & 19.5   & 95.1     & -3.6  & 135.6   & 296.7 & 2867.2   & -18.0  & 279.7    \\
                 10    & 0.5 & 76.7   & 666.6    & -4.5  & 145.1   & 136.8 & 1267.7   & 10.5   & 5.0      \\
                     & 0.7 & -63.5  & 735.3    & 4.2   & 57.7    & 48.5  & 385.4    & 30.3   & 202.8    \\
                     & 0.9 & -16.4  & 263.6    & 42.9  & 329.0   & 40.7  & 306.5    & 102.1  & 921.1\\\hline
	\end{tabular}
	\label{TabConfCu}
\end{table}
 
\begin{table}[htbp!]
    \centering
    \caption{Mean errors for simulation with no confounding and cubic non-linear  relationship.}
    \begin{tabular}{cccc}
         \textbf{Naive}& \textbf{LM} & \textbf{Lasso} & \textbf{RF}  \\ \hline
         10074 & 5747 & 38507 & 11827 \\ \hline
    \end{tabular}
     \label{TabErrorConfCu}
\end{table}

\newpage
\subsection{Confounding; linear relationship}

The results from the single feature confounding with a linear relationship, shown in Table \ref{TabConfLin},  were quite different from those of the no confounding simulation.  We first note that naive did not match $ATE_{true}$ for any of the combinations, but included a bias, likely from the confounding.  Relatively large errors were computed for naive, with average error of 812\% (see Table \ref{TabErrorConfLin}.  This shows that simply using $ATE_{naive}$ to compute treatment effects is not advised if confounding bias exists in the data.

Second, the machine learning models show mixed performance, with the linear regression model again performing exceptionally well with an average error of 0.33\% over all simulations.  Lasso performed relatively well with 17.4\% average error (less than 20\%) but random forest performed poorly with an average error of 265\%. It is surprising that RF performed poorly when it has the inherent ability to work well with non-linear data.

Again, for all models and all $ATE_{true}$ and $\pi$ values, the errors for $ATE_{true}$ = 0.1 are orders of magnitude higher than for other $ATE_{true}$ values.  Again, this can be attributed to dividing by a small fraction.  

These results suggest that for a dataset with a linear relationship and a single feature confounder, it is best to use linear regression to compute ATE using the counterfactual prediction method.

\begin{table}[htbp!]
	\centering
	\caption{$ATE_{sim}$ results for single feature confounding, linear simulation.}
	\begin{tabular}{cccccccccc}
		$ATE_{true}$ & $\pi$ & Naive  & E(Naive) & LM & E(LM) & Lasso & E(Lasso)  & RF & E(RF)  \\ \hline
			 & 0.1 & -13.84 & 38.36  & -10.00 & 0.01  & -10.28 & 2.83   & -11.25 & 12.49  \\
                     & 0.3 & -13.83 & 38.33  & -10.00 & 0.01  & -10.28 & 2.81   & -11.25 & 12.52  \\
                   -10  & 0.5 & -13.84 & 38.36  & -10.00 & 0.02  & -10.28 & 2.83   & -11.26 & 12.61  \\
                     & 0.7 & -13.84 & 38.39  & -10.00 & 0.01  & -10.28 & 2.84   & -11.26 & 12.57  \\
                     & 0.9 & -13.84 & 38.38  & -10.01 & 0.06  & -10.29 & 2.88   & -11.26 & 12.61  \\ \hline
 & 0.1 & -8.83  & 76.57  & -5.00  & 0.02  & -5.18  & 3.66   & -6.26  & 25.20  \\
                     & 0.3 & -8.83  & 76.67  & -5.00  & 0.10  & -5.18  & 3.70   & -6.26  & 25.12  \\
                    -5 & 0.5 & -8.84  & 76.75  & -5.00  & 0.04  & -5.18  & 3.65   & -6.26  & 25.16  \\
                     & 0.7 & -8.83  & 76.69  & -5.00  & 0.08  & -5.18  & 3.68   & -6.26  & 25.20  \\
                     & 0.9 & -8.84  & 76.74  & -5.00  & 0.09 & -5.18  & 3.55   & -6.25  & 24.94  \\ \hline
 & 0.1 & -3.74  & 3836  & 0.10   & 0.17 & 0.02   & 79.22 & -1.15  & 1253  \\
                     & 0.3 & -3.73  & 3829  & 0.10   & 0.58  & 0.02   & 79.08 & -1.15  & 1254  \\
                   0.1  & 0.5 & -3.74  & 3841  & 0.10   & 0.77  & 0.02   & 79.70 & -1.15  & 1254  \\
                     & 0.7 & -3.73  & 3830  & 0.10   & 3.19  & 0.02   & 75.32 & -1.15  & 1248  \\
                     & 0.9 & -3.74  & 3837  & 0.10   & 2.78 & 0.02   & 81.02 & -1.16  & 1258  \\ \hline
  & 0.1 & 1.16   & 76.80 & 5.00   & 0.00  & 5.02   & 0.41   & 3.75   & 25.10 \\
                     & 0.3 & 1.19   & 76.30 & 5.00   & 0.01  & 5.02   & 0.46   & 3.75   & 24.98 \\
                   5  & 0.5 & 1.16   & 76.80 & 5.00   & 0.03 & 5.02   & 0.40   & 3.74   & 25.19 \\
                     & 0.7 & 1.16   & 76.71 & 5.00   & 0.08 & 5.02   & 0.32   & 3.75   & 25.09 \\
                     & 0.9 & 1.16   & 76.89 & 5.00   & 0.08 & 5.01   & 0.30   & 3.75   & 25.10 \\ \hline
 & 0.1 & 6.17   & 38.25 & 10.00  & 0.01 & 10.12  & 1.22   & 8.74   & 12.55 \\
                     & 0.3 & 6.16   & 38.37 & 10.00  & 0.03 & 10.12  & 1.20   & 8.74   & 12.56 \\
                  10   & 0.5 & 6.17   & 38.31 & 10.00  & 0.02  & 10.13  & 1.27   & 8.75   & 12.53 \\
                     & 0.7 & 6.17   & 38.35 & 10.00  & 0.05  & 10.13  & 1.27   & 8.75   & 12.52 \\
                     & 0.9 & 6.17   & 38.28 & 10.00  & 0.00  & 10.12  & 1.23   & 8.75   & 12.51 \\ \hline
	\end{tabular}
		\label{TabConfLin}
\end{table}

\begin{table}[htbp!]
    \centering
    \caption{Mean errors for simulation with confounding and linear  relationship.}
    \begin{tabular}{cccc}
         \textbf{Naive}& \textbf{LM} & \textbf{Lasso} & \textbf{RF}  \\ \hline
         812 & 0.33 & 17.4 & 265 \\ \hline
    \end{tabular}
     \label{TabErrorConfLin}
\end{table}

\subsection{Confounding; non-linear (squared)}
The results for the confounding and squared non-linear simulation are shown in Tables \ref{TabConfSq} and \ref{TabErrorConfSq}.  We see that the linear regression model (LM), with an average error of 11.24\%, again outperforms all other methods when non-linearity is introduced.  The lasso model is now shown to completely blow up with an average error of over 11,000\%.   Although random forest (RF) performs poorly (781\%) it performs better than lasso, probably due to being able to handle non-linearities better than lasso.

\begin{table}[htbp!]
	\centering
	\caption{$ATE_{sim}$ results for single feature confounding and squared non-linear simulation.}
	\begin{tabular}{cccccccccc}
			$ATE_{true}$ & $\pi$ & Naive  & E(Naive) & LM & E(LM) & Lasso & E(Lasso)  & RF & E(RF)  \\ \hline
		 & 0.1 & -13.06 & 30.58  & -10.00 & 0.02   & -65.54 & 555.35  & -13.62 & 36.24 \\
                     & 0.3 & -15.41 & 54.09  & -10.08 & 0.78   & -68.69 & 586.91  & -13.74 & 37.41 \\
                  -10   & 0.5 & -14.14 & 41.39  & -9.98  & 0.16   & -67.57 & 575.69  & -13.70 & 36.99 \\
                     & 0.7 & -14.12 & 41.17  & -9.96  & 0.44   & -66.31 & 563.09  & -13.71 & 37.07 \\
                     & 0.9 & -13.91 & 39.11  & -10.03 & 0.30   & -66.31 & 563.09  & -13.67 & 36.70 \\ \hline
 & 0.1 & -8.81  & 76.10  & -5.00  & 0.08   & -61.97 & 1139.36 & -8.69  & 73.81 \\
                     & 0.3 & -8.74  & 74.75  & -4.95  & 0.99   & -62.00 & 1140.03 & -8.64  & 72.70 \\
                   -5  & 0.5 & -8.81  & 76.29  & -5.00  & 0.03   & -62.79 & 1155.78 & -8.64  & 72.90 \\
                     & 0.7 & -9.04  & 80.83  & -4.99  & 0.19   & -62.97 & 1159.34 & -8.67  & 73.32 \\
                     & 0.9 & -9.60  & 91.92  & -5.10  & 1.99   & -62.41 & 1148.17 & -8.66  & 73.25 \\ \hline
 & 0.1 & -3.95  & 4046   & 0.14   & 42.59  & -55.91 & 56010   & -3.62  & 3718  \\
                     & 0.3 & -2.15  & 2251   & -0.03  & 128.46 & -55.46 & 55560   & -3.54  & 3644  \\
                 0.1    & 0.5 & -3.52  & 3620   & 0.14   & 42.98  & -56.64 & 56745   & -3.62  & 3720  \\
                     & 0.7 & -4.95  & 5045   & 0.12   & 18.31  & -56.66 & 56764   & -3.56  & 3661  \\
                     & 0.9 & -3.48  & 3584   & 0.14   & 35.87  & -57.04 & 57143   & -3.59  & 3693  \\ \hline
   & 0.1 & -0.55  & 110.93 & 5.04   & 0.76   & -51.66 & 1133.28 & 1.36   & 72.85 \\
                     & 0.3 & 1.52   & 69.68  & 5.09   & 1.86   & -51.40 & 1127.97 & 1.29   & 74.17 \\
                  5   & 0.5 & 0.93   & 81.48  & 5.10   & 1.98   & -52.37 & 1147.44 & 1.30   & 74.06 \\
                     & 0.7 & 3.33   & 33.45  & 4.97   & 0.52   & -51.74 & 1134.76 & 1.33   & 73.31 \\
                     & 0.9 & 0.86   & 82.80  & 4.98   & 0.36   & -52.59 & 1151.78 & 1.31   & 73.72 \\ \hline
  & 0.1 & 6.30   & 37.04  & 9.98   & 0.23   & -46.01 & 560.13  & 6.36   & 36.39 \\
                     & 0.3 & 5.91   & 40.88  & 9.98   & 0.22   & -45.58 & 555.77  & 6.31   & 36.85 \\
                 10    & 0.5 & 5.73   & 42.67  & 9.96   & 0.40   & -47.41 & 574.14  & 6.31   & 36.88 \\
                     & 0.7 & 7.71   & 22.90  & 10.08  & 0.77   & -46.27 & 562.68  & 6.31   & 36.93 \\
                     & 0.9 & 5.41   & 45.92  & 9.92   & 0.84   & -46.71 & 567.10  & 6.29   & 37.07 \\\hline
	\end{tabular}
	\label{TabConfSq}
\end{table}
 
\begin{table}[htbp!]
    \centering
    \caption{Mean errors for simulation with confounding and squared non-linear  relationship.}
    \begin{tabular}{cccc}
         \textbf{Naive}& \textbf{LM} & \textbf{Lasso} & \textbf{RF}  \\ \hline
         788 & 11.24 & 11972 & 781 \\ \hline
    \end{tabular}
     \label{TabErrorConfSq}
\end{table}

\newpage
\subsection{Confounding, non-linear (cubed)}
The results for the confounding, cubic non-linear simulation are shown in Table \ref{TabConfCube} and the average errors are shown in Table \ref{TabErrorA11}.  The general results clearly show that as the relationship between $x_3$ and the output become highly non-linear, the counterfactual prediction method performs increasingly worse with unacceptably high errors.  The linear model (1614\%) and random forest (1180\%) are shown to perform better than the lasso model and naive method.  The lasso model had egregious results with errors of more than 50,000\% (for $ATE_{true}$ not equal to 0.1) and an average error of over 1,000,000\%.  The random forest performed the best, most likely attributed to being able to handle non-linear data better than linear regression models. 

\begin{table}[htbp!]
	\centering
	\caption{$ATE_{sim}$ results for single feature confounding and cubic non-linear simulation.}
	\begin{tabular}{cccccccccc}
		$ATE_{true}$ & $\pi$ & Naive  & E(Naive) & LM & E(LM) & Lasso & E(Lasso)  & RF & E(RF)  \\ \hline
	 & 0.1 & 68.43   & 784.33  & -12.11 & 21.12  & -5949 & 59387   & -12.61 & 26.06  \\
                     & 0.3 & 24.08   & 340.76  & -6.67  & 33.30  & -6043 & 60331   & -15.44 & 54.36  \\
                  -10   & 0.5 & 42.06   & 520.57  & 8.01   & 180.12 & -6028 & 60178   & -17.70 & 76.96  \\
                     & 0.7 & -26.81  & 168.09  & -19.40 & 93.97  & -5918 & 59085   & -13.03 & 30.32  \\
                     & 0.9 & -8.03   & 19.67   & -0.90  & 91.01  & -5799 & 57894   & -13.69 & 36.95  \\ \hline
  & 0.1 & 72.92   & 1558    & -8.30  & 65.94  & -5871 & 117315  & -9.44  & 88.74  \\
                     & 0.3 & -82.20  & 1544    & -15.18 & 203.59 & -6027 & 120440  & -7.25  & 45.09  \\
                   -5  & 0.5 & -53.60  & 972.08  & -3.62  & 27.58  & -5891 & 117712  & -11.81 & 136.24 \\
                     & 0.7 & -40.87  & 717.35  & -20.36 & 307.11 & -5879 & 117489  & -10.01 & 100.17 \\
                     & 0.9 & 21.32   & 526.40  & -6.43  & 28.70  & -6087 & 121644  & -8.03  & 60.51  \\ \hline
 & 0.1 & -20.42  & 20519   & -0.68  & 782.05 & -5913 & 5912943 & -2.07  & 2168   \\
                     & 0.3 & -5.92   & 6023    & -7.37  & 7466   & -5911 & 5911430 & -5.01  & 5111   \\
                   0.1  & 0.5 & 63.28   & 63182   & -16.87 & 16973  & -5868 & 5868095 & -7.36  & 7462   \\
                     & 0.7 & -113.81 & 113906  & 5.42   & 5320   & -5861 & 5861042 & -2.72  & 2816   \\
                     & 0.9 & -29.26  & 29362   & -7.73  & 7830   & -6004 & 6004508 & -10.55 & 10651  \\ \hline
   & 0.1 & -12.03  & 340.68  & -2.67  & 153.42 & -5907 & 118250  & 1.90   & 61.96  \\
                     & 0.3 & 28.19   & 463.72  & 0.06   & 98.84  & -6009 & 120276  & 1.39   & 72.21  \\
                   5  & 0.5 & -57.60  & 1251.98 & 27.29  & 445.83 & -5948 & 119066  & 3.35   & 33.05  \\
                     & 0.7 & 67.14   & 1242.81 & 4.87   & 2.67   & -5892 & 117943  & -2.70  & 154.05 \\
                     & 0.9 & 33.74   & 574.82  & 6.33   & 26.58  & -5896 & 118028  & -0.65  & 112.92 \\ \hline
  & 0.1 & 27.38   & 173.83  & 6.41   & 35.88  & -6021 & 60313   & 2.55   & 74.54  \\
                     & 0.3 & -17.86  & 278.65  & 11.56  & 15.62  & -6101 & 61112   & 3.00   & 70.02  \\
                   10  & 0.5 & -1.67   & 116.70  & -1.68  & 116.81 & -5978 & 59883   & 5.89   & 41.05  \\
                     & 0.7 & -8.66   & 186.62  & 12.70  & 27.02  & -5993 & 60034   & 7.53   & 24.69  \\
                     & 0.9 & 22.42   & 124.17  & 7.62   & 23.83  & -5821 & 58308   & 8.48   & 15.21 \\\hline
	\end{tabular}
	\label{TabConfCube}
\end{table}

\begin{table}[htbp!]
    \centering
    \caption{Mean errors for simulation with confounding and cubic non-linear  relationship.}
    \begin{tabular}{cccc}
         \textbf{Naive}& \textbf{LM} & \textbf{Lasso} & \textbf{RF}  \\ \hline
         9795 & 1614 & 1253708 & 1180 \\ \hline
    \end{tabular}
     \label{TabErrorA11}
\end{table}

\subsection{Removing \texorpdfstring{$ATE_{true}$} == 0.1 }
Due to the instability of the errors for $ATE_{true}$ = 0.1, we recalculated the errors with $ATE_{true}$ = 0.1 removed.  The results for no confounding and confounding are presented in Tables \ref{TabErrorA1} and \ref{TabErrorA2} respectively. 

It can now be seen that for no confounding (Table \ref{TabErrorA1}), the machine learning method performs satisfactorily \textit{even with squared non-linearity}.  The largest errors for non-linear squared and no confounding are 26.2\% with lasso model.  Again, this method does not work with cubic non-linearity.   

  For confounding results (\ref{TabErrorA2}) and linear data, the machine learning models perform satisfactorily with a largest error of 18.8\%.  However, for non-linear squared, we still compute excellent results for LM but not for any other models.  
  
  To summarize, good results are found for both no confounding and confounding, as long as the data is linear; and acceptable results can be achieved for squared non-linearity for both no confounding and confounding, depending on the type of model used.  Results for cubed non-linear are unacceptable.

\begin{table}[htbp!]
    \centering
    \caption{Mean errors for no confounding simulations, excluding results from $ATE_{true}$ = 0.1. }
    \begin{tabular}{lllll}
         &\textbf{Naive}& \textbf{LM} & \textbf{Lasso} & \textbf{RF}  \\ \hline
         Linear&0.16 & 0.03 & 0.15 & 0.14 \\ 
        Non-linear (Squared) &8.60 & 2.50 & 26.2 & 5.03 \\ 
         Non-linear (Cubed)& 804.7 & 393.9 & 2784 & 780.3 \\ \hline

    \end{tabular}
     \label{TabErrorA1}
\end{table}

\begin{table}[htbp!]
    \centering
    \caption{Mean errors for  confounding simulations, excluding results from $ATE_{true}$ = 0.1. }
    \begin{tabular}{lllll}
         &\textbf{Naive}& \textbf{LM} & \textbf{Lasso} & \textbf{RF}  \\ \hline
         Linear&57.5 & 0.04 & 2.00 & 18.8 \\ 
        Non-linear (Squared) &58.7 & 0.65 & 855 & 55.1 \\
         Non-linear (Cubed)& 595.3 & 99.9 & 89234 & 65.7 \\ \hline

    \end{tabular}
     \label{TabErrorA2}
\end{table}

\section{Discussion}
\label{Sec_discussion}
This study presents, as far as we know, the first simulation study investigating the accuracy of the counterfactual prediction method presented in the literature \cite{Beemer2017, Beemer2018, BevanSmith2020}.  In this study we were able to measure this method against the true treatment effects.

As expected, the results suggest that the naive method should never be used to compute treatment effects in observational studies, unless we are certain that there is no confounding \textit{and} the data does not have excessive non-linearity.  It should never be used when confounding is present.

Perhaps the main finding is that non-linearity plays a vital role in the success of this counterfactual prediction method.  The degree of non-linearity may play an even more important role than whether confounding is present or not.   Even when there is no confounding, errors are egregiously  high for both squared and cubed non-linearity.  When there \textit{is} confounding, as long as the data is linear, models might be able to perform well (LM =0.33\%) or satisfactorily (lasso = 17.4\%).  

It is most surprising that the linear regression model outperforms lasso and RF, whether confounding is present or not.  It was expected that the lasso model would outperform the linear model due to its' inherent ability to deal with multi-collinearity; which is essentially what confounding is.   
The use of regularized regression, such as lasso or ridge regression, to adjust for confounding, has been used before in the literature \cite{Franklin2015,Greenland2008}.  Franklin et al. \cite{Franklin2015} compared ridge and lasso regression with high-dimensional propensity score estimations to adjust for confounding in a simulation study.  The benefit of using regularized regression is that it could potentially deal with confounding by addressing multi-collinearity between the confounder (\textit{X}) and the treatment (\textit{T}).  However, even though lasso performed well for both no confounding and confounding linear simulations, the linear regression model still outperformed lasso.   As soon as non-linearity was introduced, the lasso model performs poorly. The lasso is based upon the linear regression model, but uses a more restrictive method for estimating the coefficients (as discussed in Section \ref{sec5}).  Lasso can therefore be less flexible than linear regression \cite{Hastie2009} and may be why it does not outperform simple linear regression.  

In the presence of confounding and cubic non-linearity, the results show that random forest begins to outperform the other models, although still performing very poorly (65.7\% from Table \ref{TabErrorA2}).  This is most likely due to being able to handle non-linearity well \cite{Hastie2009}. When training the random forest models, we aimed to tune the hyperparameters as well as possible; but tuning of these parameters is often a challenging task.  In this study, we did not make use of exhaustive grid search or random grid search methods with a wide range of random forest hyperparameters. Tuning in this way could potentially improve RF performance and reduce the absolute errors.  This is suggested for future work.   

The results of this simulation suggest that as long as the data is linear, whether confounding exists or not, the machine learning counterfactual prediction method works satisfactorily, if not very well, depending on the type of model used.  As soon as non-linearity is introduced into the data, this method tends to break down.

\subsection{Limitations and recommendations for future work}
This section discusses the limitations of this study and how it presents opportunities for further work.

  First, the confounding simulation was based on a single observable confounding feature affecting both the treatment and output.  We did not look at multiple confounding features. It is predicted that more confounding would introduce larger errors. Future work can look at more observable confounders.

Second, we did not model \textit{hidden} confounding which is common in observational studies. The assumption was no hidden confounding.  Future work could study the results of simulations that include hidden confounding.

Third, more research is needed to understand how different models such as neural networks or boosting will perform.  More thorough tuning of model hyperparameters is suggested for more complex models such as neural networks, random forest and boosting models.

Fourth, in this study we made use of only 5 input features.  Further study is required for datasets with a larger number of input features.  

Finally, when estimating treatment effects using model predictions, we always used the same model on both treated and control groups.  For example, we used linear regression on both groups and estimated treatment effects.  Or we used random forest on both groups and estimated treatment effects. Future work could look at mixing up the models.  For example, using a lasso model on a treated group and a random forest on the control and then estimating treatment effects.

\bibliographystyle{unsrt}  
\bibliography{bibliography.bib}

\end{document}